\documentclass[final, 5p, times, twocolumn, sort&compress]{elsarticle}

\usepackage[utf8]{inputenc}
\usepackage{CJKutf8}
\usepackage{amssymb}
\usepackage{amsmath,amssymb}
\usepackage{autobreak}
\usepackage{algpseudocode}
\usepackage{algorithm}
\usepackage{url}
\usepackage{graphicx}
\usepackage{float}
\usepackage{comment}
\usepackage{bm}
\newcommand{\vect}[1]{\boldsymbol{\mathbf{#1}}}
\def\vec#1{\vect{#1}}
\usepackage[switch]{lineno}
\usepackage{color}
\allowdisplaybreaks[1]

\newcommand\HL[1]{{\color{black}#1}}

\newcommand\HLL[1]{{\color{black}#1}}

\newcommand\HLLL[1]{{\color{black}#1}}

\newcommand\HLT[1]{{\color{black}#1}}

\journal{journal}

\begin{document}

\begin{frontmatter}

\title{Effect of congestion avoidance due to congestion information provision on optimizing agent dynamics on an endogenous star network topology}

\author[UTOKYOA]{\textasteriskcentered Satori Tsuzuki}
\ead{tsuzukisatori@g.ecc.u-tokyo.ac.jp}

\author[UTOKYOA,UTOKYOB,UTOKYOC]{Daichi Yanagisawa}
\ead{tDaichi@mail.ecc.u-tokyo.ac.jp}

\author[UTOKYOA,UTOKYOB,UTOKYOC]{Katsuhiro Nishinari}
\ead{tknishi@mail.ecc.u-tokyo.ac.jp}

\address[UTOKYOA]{Research Center for Advanced Science and Technology, The University of Tokyo, 4-6-1, Komaba, Meguro-ku, Tokyo 153-8904, Japan}
\address[UTOKYOB]{Department of Aeronautics and Astronautics, School of Engineering, The University of Tokyo, 7-3-1, Hongo, Bunkyo-ku, Tokyo, 113-8656, Japan}
\address[UTOKYOC]{Mobility Innovation Collaborative Research Organization, The University of Tokyo, 5-1-5, Kashiwanoha, Kashiwa-shi, Chiba, 277-8574, Japan}

\begin{abstract}
The importance of fundamental research on network topologies is widely acknowledged. This study aims to elucidate the effect of congestion avoidance of agents given congestion information on optimizing traffic in a network topology. We investigated stochastic traffic networks in a star topology with a central node connected to isolated secondary nodes with different preferences. Each agent at the central node selects a secondary node by referring to the declining preferences based on the congestion rate of the secondary nodes. We examined two scenarios: 1) Each agent can repeatedly visit the central and secondary nodes. 2) Each agent can access each secondary node only once. For 1), we investigated the uniformity of the agent distribution in a stationary state, and for 2), we measured the travel time for all agents visiting all nodes. When agents repeatedly visit central and other nodes, the uniformity of agent distribution has been found to show three types of nonlinear dependence on the increase in nodes. We found that multivariate statistics describe these characteristic dependences well, suggesting that the balance between the equalization of network usage by avoiding congestion and the covariance caused by mutual referral to congestion information determines the uniformity. We discovered that congestion-avoidance linearizes the travel time, which increases exponentially with the number of nodes, notwithstanding the degree of reference to the congestion information. Consequently, we successfully described the optimization effect of congestion-avoidance on the collective dynamics of agents in star topologies. Our findings are useful in many areas of network science.
\end{abstract}


\end{frontmatter}







\section{Introduction}
Network science has been acknowledged as a crosscutting discipline that pursues the universal nature of network characteristics. Fundamental research on network topologies has been particularly significant for many applications ~\cite{doi:10.1098/rsta.2012.0375}. Namely, the knowledge in network topologies is beneficial not only for a specific field but also for various areas in science, engineering, and technologies. Examples include the Internet, telecommunications, ecological systems, and transport networks~\cite{10.1145/2699416, Ings2018, dale_fortin_2021}. It is necessary to analyze structural properties, such as the average path lengths or node degrees~\cite{Wahlisch2010, PhysRevE.81.046104, PhysRevE.88.012812}. Furthermore, there is a strong demand to explore transport properties, which are the distribution of agents and their traffic in topological networks, where agents usually represent creatures, humans, or objects. Graph theory~\cite{Lesne2006, Diestel2017} and queuing theory~\cite{ERLANG-A-K, https://doi.org/10.1002/net.3230060210} are effective tools for the theoretical analysis of traffic transportation~\cite{doi:https://doi.org/10.1002/9781118625651.ch2, 10.1007/978-3-540-79992-4_59, YANAGISAWA2013238, PhysRevE.98.042102}. However, the applications of these theories to complex real-world transport networks are limited. As a remedial measure, cellular automata~\cite{10.5555/1102024, RevModPhys.55.601} or multi-agent simulations have been utilized as supportive but powerful approaches to complex network problems~\cite{YAMAMOTO2007654, CDA14, Tsuzuki_2020}.

Elucidating the effect of agents' congestion avoidance on their collective dynamics in basic topologies can help deepen the understanding of the mechanism of ``traffic jams'' in various network sciences. However, such an effect has not been fully discussed in multivariate statistics. In studies of traffic transport networks, there have been two different approaches to analyzing network problems: conceptual and semi-empirical. The former examines the basic topologies or conceptual network models to grasp the essence, whereas the latter approach develops generic models and adapts them to complicated phenomena by optimizing parameters according to experimental data. 
Many related studies have primarily focused on applications in engineering; thus, they used semi-empirical approaches~\cite{Balouchzahi2019, MA2016164, CHEN2017229, CHEN2015339}. 
For example, several studies have parameterized multiple types of traffic information [e.g., incidents, congestion, and travel time] and calculated cost-minimization functions using optimization algorithms to improve traffic efficiency~\cite{Balouchzahi2019, MA2016164}. One study used traffic information as the cost of searching the shortest routes~\cite{CHEN2017229} using the A-star algorithm~\cite{4082128}. Other studies applied the Bayesian network~\cite{CHEN2015339} or logit regressions~\cite{Zhang2014} to a set of traffic information to determine the relationship between travel time and drivers' preferences on real-world road networks. Briefly, many previous studies can be largely classified into three methodologies according to their application purposes: optimizations, graph theory, and regressions. 

Several studies have applied multivariate statistics to traffic transport networks~\cite{STATHOPOULOS2003121, YAN2017149, NAKAYAMA2016238}. \HL{Notably}, the literature \cite{NAKAYAMA2016238} presented an optimization of travel time by using traffic information provision. They focused on a general stochastic network that connects multiple origin-destination pairs, calculated the variance-covariance matrix of the route flow in the system, and utilized it with a network equilibrium model~\cite{ARNOTT1991309, CHEN2010493, Huang2011, doi:10.1287/trsc.1110.0357}. They demonstrated the shortening of the travel time in a road network in a city by using the stationary conditions of stochastic network equilibrium. 
\HL{Although} the literature' s study \cite{NAKAYAMA2016238} is impressive because it \HL{designed a} multivariate \HL{statistical model} of agent dynamics with traffic information provision, it assumed several constraints for application purposes. \HL{In particular}, agents decide on their destinations only due to exogenous factors. In other words, traffic information is given to agents deterministically, and the resulting route choices of agents do not provide feedback into the input traffic information. This condition is required for the system to satisfy the multivariate normal distribution (mnd)~\cite{Tong1990, Flury1997chap2, Flury1997chap3, ROUSSAS2014179}, which is indispensable for their optimization using stochastic network equilibrium. 
Unfortunately, real-world traffic networks are primarily endogenous. Congestion information is stochastically given to agents, and the resulting congestion status is often fed back into the input congestion information. It is known that the relationship between stochastic input variables and output distribution, that is, endogenous or exogenous of the network, seriously affects system dynamics. Because of these complexities, to the best of our knowledge, no study has fully discussed the effects of providing congestion information on the fundamental characteristics of endogenous traffic networks in terms of multivariate statistics.

This study aims to clarify the effect of agents' congestion avoidance due to congestion information provision on traffic properties in a network topology. We focused on a star topology, in which a central primary node is connected to multiple secondary nodes that are isolated from each other. 
\HLT{We can find star topologies in various fields. A star topology is one of seven basic networks and analyzing the flow of data in a star topology is essential to computer networks. Collective dynamics in a star topology have attracted attention in epidemiological or ecological metapopulations. In addition to these physical networks, a star topology is often used as a conceptual decision-making model by individuals with multiple choices. The convolution of numerous star topologies yields neural networks. Accordingly, clarifying the general characteristic of a star topology will also contribute to non-specific disciplines.} 
\HLT{In this study,} we investigated the dynamics of a stochastic transportation network in which each agent at the primary node stochastically chooses one of the secondary nodes by referring to the fixed access rates, that is, the preferences, declined by the congestion rate of each secondary node. Specifically, we examined the following two scenarios: (1) each agent can access the same secondary node repeatedly and (2) each agent can access each secondary node only once. We refer to the former scenario as Scenario-1 and the latter as Scenario-2.
We investigated the uniformity of agent distribution in the stationary state in Scenario-1 and measured the travel time for all agents visiting all the nodes in Scenario-2. This paper discusses the observations in terms of multivariate statistics. 

A key feature of our model is that the system stochastically provides congestion information to agents. 
The agent distribution or expected number of agents at a secondary node in a stationary state depends on fixed preferences and the congestion rate, which is \HL{in proportion} to the number of agents in each node. Accordingly, the number of agents operates as both input and output variables. The resulting agent distribution among the secondary nodes feeds back into the stochastic input variable regarding the congestion rate; the network is \HL{purely} endogenous. 
In multivariate statistics, the endogenous system has been challenging to study because it often shows irregular behaviors that do not necessarily follow standard statistical models \HL{including the mnd}. Therefore, this study investigates whether we can apply the framework of multivariate statistics to such an endogenous traffic network with a star topology. \HL{Furthermore}, we theoretically derive a variance-covariance matrix without assuming that the agent distribution follows mnd, to discuss the \HL{generic} case involving small systems.
Section~\ref{sec:targetsysmodel} provides more details of our model. The results of this conceptual study can be helpful for many related scientific areas. Scenario-1 can provide a new perspective on metapopulation networks~\cite{Limdi2018, Rao2021}. Scenario-2 reports an applied case of the traveling salesman problems~\cite{HOOS2005357}, where we consider the \HL{effect of} congestion information provision \HL{on facilitating} travel on \HL{a star} network topology.

The remainder of this paper is organized as follows. In section 2, we provide a detailed description of the target system. We also derive analytical solutions for a specific case in which agents are provided with no traffic information by solving the state-transition equations. We then report the results of small preliminary simulations in both scenarios to clarify the following two characteristics: (a) the equalization of network usage owing to congestion avoidance of agents observed in Scenario-1. (b) the shortening effect of travel time due to the equalization effect observed in Scenario-2. Considering the results of the preliminary tests, we define a criteria state indicating the uniform distribution of agents based on the above-mentioned analytical solution as a preparation for the subsequent sections. In Section 3, we investigate the uniformity of agent distributions in Scenario-1. We observed that the uniformity of agent distribution shows three types of nonlinear dependence on the increase in nodes. We also report that the congestion avoidance of agents linearizes the travel time, which exponentially increases with the number of nodes as long as agents are provided with congestion information, notwithstanding the degree of reference to the congestion information. In Section 4, we derive a theoretical model based on multivariate statistics and compare it with the simulation results, reporting that our model clearly describes the observed characteristic dependence in Scenario-1. In particular, our analysis indicates the following: the balance between the equalization of network usage by avoiding congestion and the covariance caused by mutually referring to congestion information determines overall uniformity. We also discuss why the travel time step exponentially increases when agents are provided with no congestion information and are linearized otherwise. Section 5 summarizes the results and concludes the study.  

\begin{figure}[t]
\vspace{-16.0cm}
\includegraphics[width=1.0\textwidth, clip, bb= 4 4 1091 1740]{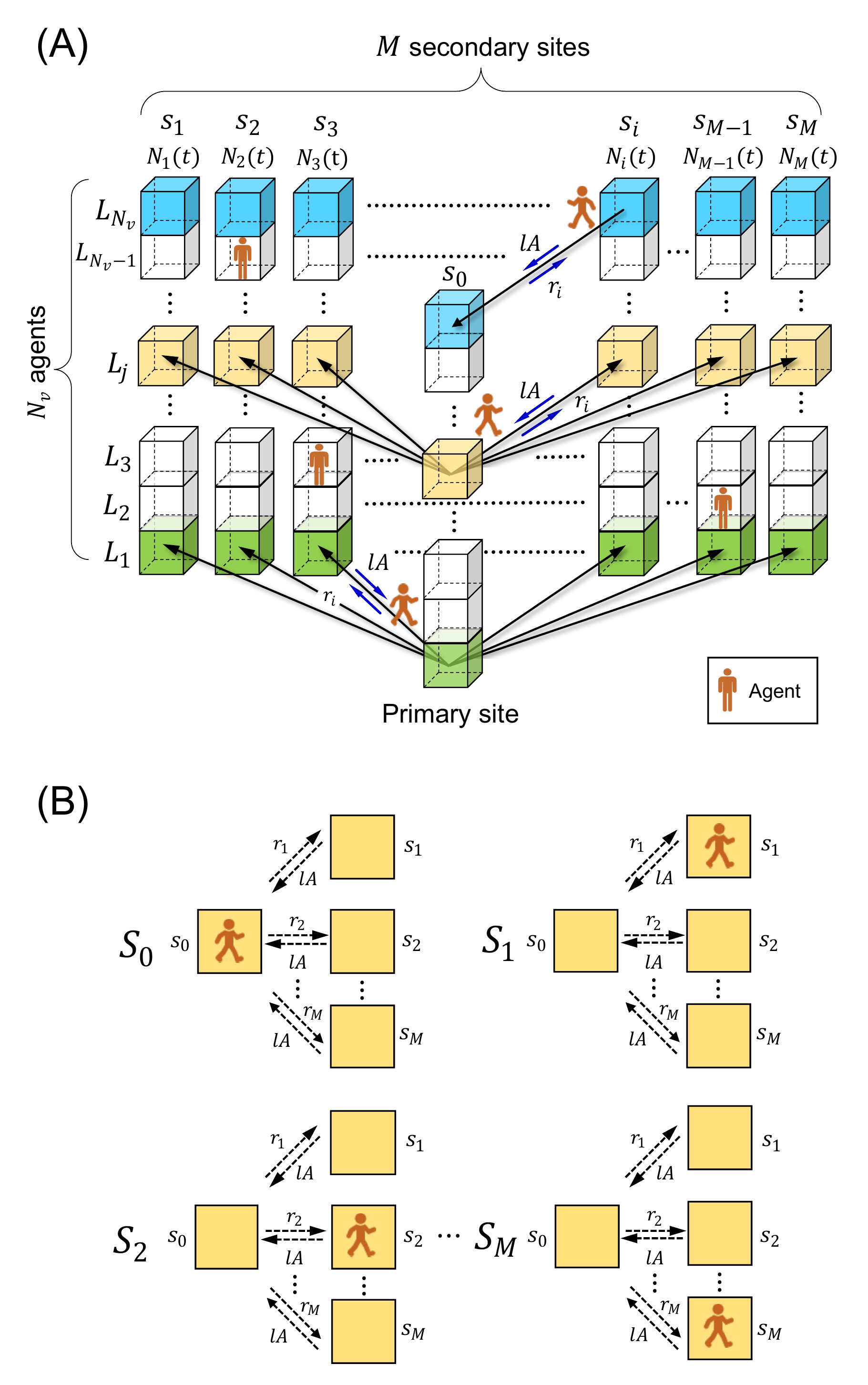}
\caption{Schematic of the target system}
\label{fig:Figure_SchemView}
\end{figure}
\section{Methods}\label{sec:targetsys}
\subsection{Model}\label{sec:targetsysmodel}
Consider a star topology with a primary node connected to $M$ isolated secondary nodes with different preferences. We denote the total number of agents and the number of agents of the $i$th secondary node at the $t$ \HL{discrete-}time step and the fixed preference of the $i$th secondary node as $N_v$, $N_i(t)$, and $w_i$, respectively. For ease of understanding, we refer to $w_{i}$ as a weight or the $i$th preference. Each agent at the primary node stochastically chooses one of the secondary nodes by referring to the preferences declined by the congestion rate of each secondary node. Accordingly, the probability that an agent moves from the primary node to the $i$th secondary node is expressed as 
\begin{eqnarray}
r_{i}&:=&\frac{A}{\xi} \cdot \Biggl\{{w}_{i} + \alpha \biggl( 1 - \frac{N_{i}(t)}{N_{v}}\biggr)\Biggr\}, \nonumber\\
	 \xi&=&\sum_{i=1}^{M} \Biggl\{ {w}_{i} + \alpha \biggl( 1 - \frac{N_{i}(t)}{N_{v}}\biggr) \Biggr\}, \label{eq:considerconges}
\end{eqnarray}
where the parameter $A$ determines the moving rate from the primary node to either of the secondary nodes and controls the outflow of agents among $N_{v}$ agents from the primary to secondary nodes. By contrast, we set the return rate from the secondary nodes to the primary node as the $l$ multiple of $A$; in other words, parameter $l$ controls the inflow from the secondary nodes to the primary node. Parameter $\alpha$ determines the effect of avoiding congestion on the decline in preference $w_i$.
Figure~\ref{fig:Figure_SchemView}(A) shows the schematic of the target system from the viewpoint of decision-making; Each layer in the horizontal direction represents all possible choices in decision-making by an identical agent. For instance, the yellow $L_j$th layer has $M$ secondary nodes that the $j$th agent can select. $s_{i}$ indicates the primary node when $i~=~0$, and otherwise indicates the $i$th secondary node. Figure~\ref{fig:Figure_SchemView}(B) depicts all the possible states that an agent can take, corresponding to the yellow layer in the left part.

In the target system, each agent at the primary node stochastically selects one of the secondary nodes by referring to the preferences declined by the congestion rate of each secondary node. We examined the following two scenarios: (1) each agent can access the same secondary node repeatedly and (2) each agent can access each secondary node only once. We refer to the former scenario as Scenario-1 and the latter as Scenario-2. 
We investigated the uniformity of agent distribution in the stationary state in Scenario-1 and measured the travel time for all agents visiting all the nodes in Scenario-2. 
\HLL{In Scenario-1, the average outflow of the number of agents moving from the primary to secondary nodes is controlled by parameter $A$. By contrast, in Scenario-2, the total probability is normalized among all the secondary nodes, and the agents leave out the probabilities of choosing the already visited secondary nodes. If an already visited node is selected in a trial, the system discards the trial; it can be considered a call-loss system in Scenario-2. All the agents are updated simultaneously in both scenarios.
}
We introduce the following physical value $U$ to evaluate usage in each node:
\begin{eqnarray}
U &:=& \frac{U_{f}}{N_{v}},
\end{eqnarray}
where $U_{f}$ represents the number of agents in a secondary node after reaching a stationary state, and $N_{v}$ denotes the total number of agents. Regarding travel time in Scenario-2, we count the number of time steps required for all agents to complete visiting at all secondary nodes and refer to this as $T_s$.
Accordingly, we investigate the dependence of these two features $U$ and $T_{s}$ on the major parameters of the system. We then discuss these mechanisms from the perspective of multivariate statistics. The method for evaluating the uniformity of the distribution is described in Section~\ref{sec:evapophomo}.

\subsection{Agent distribution}\label{sec:targetsysfeature}
As depicted in Fig.~\ref{fig:Figure_SchemView}(B), each agent assumes one of the $M+1$ possible states. The geometric feature of the star topology leads to the following rules: the state at which an agent exists on the primary node, $S_0$, can be transitioned from one of the states between $S_0$ and $S_M$. In addition, each of the states between $S_1$ and $S_M$ only transitioned from state $S_0$. Accordingly, the state at time step $n+1$ is determined by the transitions from one of the $M+1$ states at time step $n$; therefore, the state transition equations between time steps $n+1$ and $n$ can be described as $\vec{S}^{(n+1)} = \vec{T}\vec{S}^{(n)}$, where $\vec{S}^{(n)}$ is a state vector at time step $n$, represented by $\vec{S}^{(n)} = [S^{n}_0, S^{n}_1, \cdots, S^{n}_M]^{t}$. $\vec{T}$ is a state transition matrix of size $M+1 \times M+1$. 
The element ${e}_{ij}$ of $\vec{T}$ in the $i$th column and the $j$th row is expressed as
\begin{eqnarray}
e_{ij} &=& 
\begin{cases}
~1-A & (i~=~0~,~j~=~0), \\
~lA & (i~=~0~,~1~\le~j~\le~M), \\
~1-lA & (i~=~j~=k, ~1~\le~k~\le~M), \\
~r_{i} & (1~\le~i~\le~M,~j~=~0), \\
~0 & (otherwise).
\end{cases}
\label{eq:statemat}
\end{eqnarray}
The relationship $\vec{S}^{(n+1)} \approx \vec{S}^{(n)}$ was established after reaching a stationary state. The exact expression of the state vector $\vec{S}^{(n)}$ is obtained by solving the \HL{state transition equations} as follows: 
\begin{eqnarray}
\vec{S}^{(n)} &:=&
\scalebox{1.0}{$\displaystyle
\begin{bmatrix}S_0^n\\S_1^n\\S_2^n\\\vdots\\S_M^n\end{bmatrix}= c/(l+1) \times \begin{bmatrix}1\\r_{1}/(lA)\\r_{2}/(lA)\\\vdots\\r_{M}/(lA)\end{bmatrix}
$}, \label{eq:statevecresultcasea}
\end{eqnarray}
where $c$ is the indeterminate parameter. We can see the difficulty of solving this problem; In case $\alpha \ne 0$, the system becomes endogenous; Namely, the number of agents at a stationary state is designated by the state $\vec{S}^{(n)}$ whereas $\vec{S}^{(n)}$ itself includes the number of agents $N_{i}(t)$, as confirmed by the right-hand side of Eq.~(\ref{eq:considerconges}). The agent distribution at the steady state can be obtained because of this autonomous self-optimization. 
In addition, the solution always includes an indeterminate parameter. In a wider sense, the target system can be said to be the Diophantine problem~\cite{Andreescu}. Meanwhile, in the case of $\alpha=0$, the second term in the curly braces in Eq.(\ref{eq:considerconges}) is always zero. Therefore, the system becomes exogenous, and we can obtain the exact solutions for Scenario-1 by setting $c$ to \HL{$l$} such that $\vec{S}^{(n)}$ satisfies the normalization of $|\vec{S}^{(n)}|=1$. In this case, $\vec{S}^{(n)}$ represents the probability distribution.

\begin{figure}[t]
\vspace{-10.5cm}
\includegraphics[width=1.0\textwidth, clip, bb= 4 4 1325 1357]{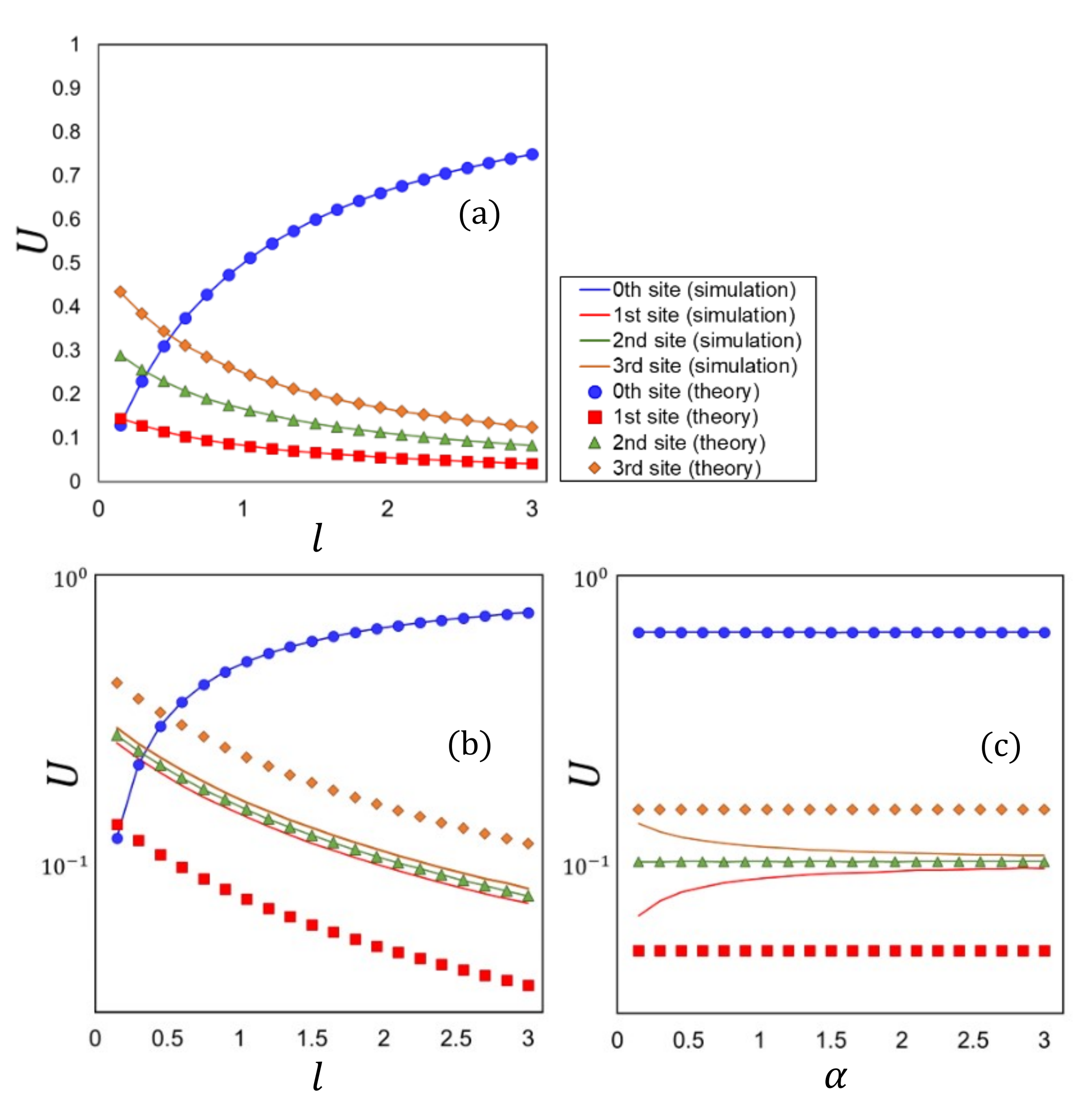}
\caption{(a) Comparison of simulations with our \HL{theoretical} model in Eq.(\ref{eq:statevecresultcasea}) when setting $\alpha = 0$ for different values of the parameter $l$ between 0 and 3.0, (b) comparison of simulations at $\alpha = 2.5$ for different values of $l$ between 0 and 3.0 with the model of Eq.(\ref{eq:statevecresultcasea}) at $\alpha = 0$, and (c) comparison of simulations at $l = 2.0$ for different values of $\alpha$ between 0 and 3.0 with the model in Eq.(\ref{eq:statevecresultcasea}) while keeping $\alpha = 0$.
}
\label{fig:Figure_ModelFeatures_FourSites}
\end{figure}
As a preliminary test for Scenario-1, we focused on a problem using $M=3$. We set the preferences ($w_{1}$, $w_{2}$, $w_{3}$) to ($1$, $2$, $3$), where the subscript $x$ of $r_{x}$ corresponds to the index of the secondary node. We set the parameters ($N_v$, $A$) to (1,000, 0.1) in this test.
\HL{We set the number of time steps for each simulation to 1,000, which was confirmed to be a sufficient number of time steps for reaching a stationary state in preliminary tests.} 
Then, we measured the usage of each secondary node for different values of $l$ and $\alpha$.
Figure~\ref{fig:Figure_ModelFeatures_FourSites}(a) shows the comparison of simulations with our theoretical model in Eq.~(\ref{eq:statevecresultcasea}) when $\alpha = 0$, for different values of the parameter $l$ between 0 and 3.0. The solid lines and circle symbols in Fig.~\ref{fig:Figure_ModelFeatures_FourSites}(a) respectively represent the simulations and model values, which agreed well in every measured range. 
As previously mentioned, we cannot obtain analytical solutions for all cases of $\alpha \ne 0$ because the second term in the curly braces of Eq.~(\ref{eq:considerconges}) includes the number of agents $N_{i}(t)$. Accordingly, we performed simulations for $\alpha = 2.5$ as an example of $\alpha \ne 0$ for different values of $l$ between 0 and 3.0 and compared them with the model values of Eq.(\ref{eq:statevecresultcasea}) with $\alpha = 0$. The results are presented in Fig.~\ref{fig:Figure_ModelFeatures_FourSites}(b) on a logarithmic scale, where the solid lines indicate the simulation results and the symbols represent the plots of the model in Eq.~(\ref{eq:statevecresultcasea}): Here, we made some interesting observations; as shown in Fig.~\ref{fig:Figure_ModelFeatures_FourSites}(b), the usage $U$ of the first node was found to approach that of the second node, which had a neutral preference. In addition, $U$ in the third node, which had the largest preference, approached the second node from the opposite direction. Namely, the usages of the three nodes were almost equal regardless of parameter $l$. 

Similarly, we performed simulations for $l=2.0$ for different values of $\alpha$ between 0 and 3.0 and compared them with the model values of Eq.~(\ref{eq:statevecresultcasea}) with $\alpha = 0$. The results are presented in Fig.~\ref{fig:Figure_ModelFeatures_FourSites}(c)). As with the results in the $l$ direction, the usage $U$ of the first node was found to approach the second node, and that of the third node approached the second node from the opposite direction as $\alpha$ increased. The increase in $\alpha$ mitigated the imbalance and facilitated the equalization of usage $U$ among nodes. 
To summarize the results of (b) and (c), we can say the following: even though the secondary nodes were set to have non-uniform preferences, the network behaved uniformly under specific conditions. In this paper, we refer to this observation as the ``equalization effect''. In Section~\ref{sec:evapophomo}, we introduce a criterion to evaluate the uniformity of the agent distribution resulting from the equalization effect.

\subsection{Evaluation of the uniformity of agent distribution}\label{sec:evapophomo}
We define the imbalance of the system as the deviation of the state vector $\vec{S}^{(n)}$ in a stationary state from an ideal state vector $\vec{S}^{(n)}_h$, where each secondary node has the same number of agents. We directly obtain the state vector $\vec{S}^{(n)}_h$ by replacing $N_{i}(t)$ in Eq.~(\ref{eq:considerconges}) with $N_{v}/M$. The total number of agents is equally divided by the number of secondary nodes, as follows:
\begin{eqnarray}
\vec{S}^{(n)}_h~~:=~~
\begin{bmatrix}S_{h, 0}^n\\S_{h, 1}^n\\S_{h, 2}^n\\\vdots\\S_{h, M}^n\end{bmatrix} &=& l/(l+1) \times \begin{bmatrix}1\\ \overline{\langle {r}_{1}\rangle}/(lA)\\ \overline{\langle {r}_{2}\rangle}/(lA)\\\vdots\\\overline{\langle {r}_{M}\rangle}/(lA)\end{bmatrix}, \nonumber \\
\overline{\langle {r}_{i}\rangle}&:=&\frac{A}{\xi} \cdot \Biggl\{{w}_{i}^{(j)} + \alpha \biggl( 1 - \frac{1}{M}\biggr)\Biggr\}, \nonumber\\
	\xi&=&\sum_{i=1}^{M} \Biggl\{ {w}_{i}^{(j)} + \alpha \biggl( 1 - \frac{1}{M}\biggr) \Biggr\}. \label{eq:statevecapprox}
\end{eqnarray}
By using Eq.~(\ref{eq:statevecresultcasea}) and Eq.~(\ref{eq:statevecapprox}), we define the imbalance ratio $I_h$ as follows: 
\begin{eqnarray}
I_h := \frac{100}{M} \sum_{i=1}^{M} \frac{\sqrt{(S_{i}^{n} - S_{h, i}^{n})^2}}{S_{h, i}^n} \label{eq:inhomo}.
\end{eqnarray}
where $S_{x}^{n}$ represents the $x$th element of the state vector $\vec{S}^{(n)}$, and $S_{h, x}^{n}$ indicates the $x$th element of an ideal state vector $\vec{S}_{h}^{(n)}$. We introduced a constant value of 100 to express $I_h$ as a percentage. 

\subsection{Travel efficiency on the network}\label{seq:sub:scenario2}
\begin{figure}[t]
\vspace{-14.6cm}
\hspace{+0.0cm}
\includegraphics[width=1.0\textwidth, clip, bb= 4 4 1027 1460]{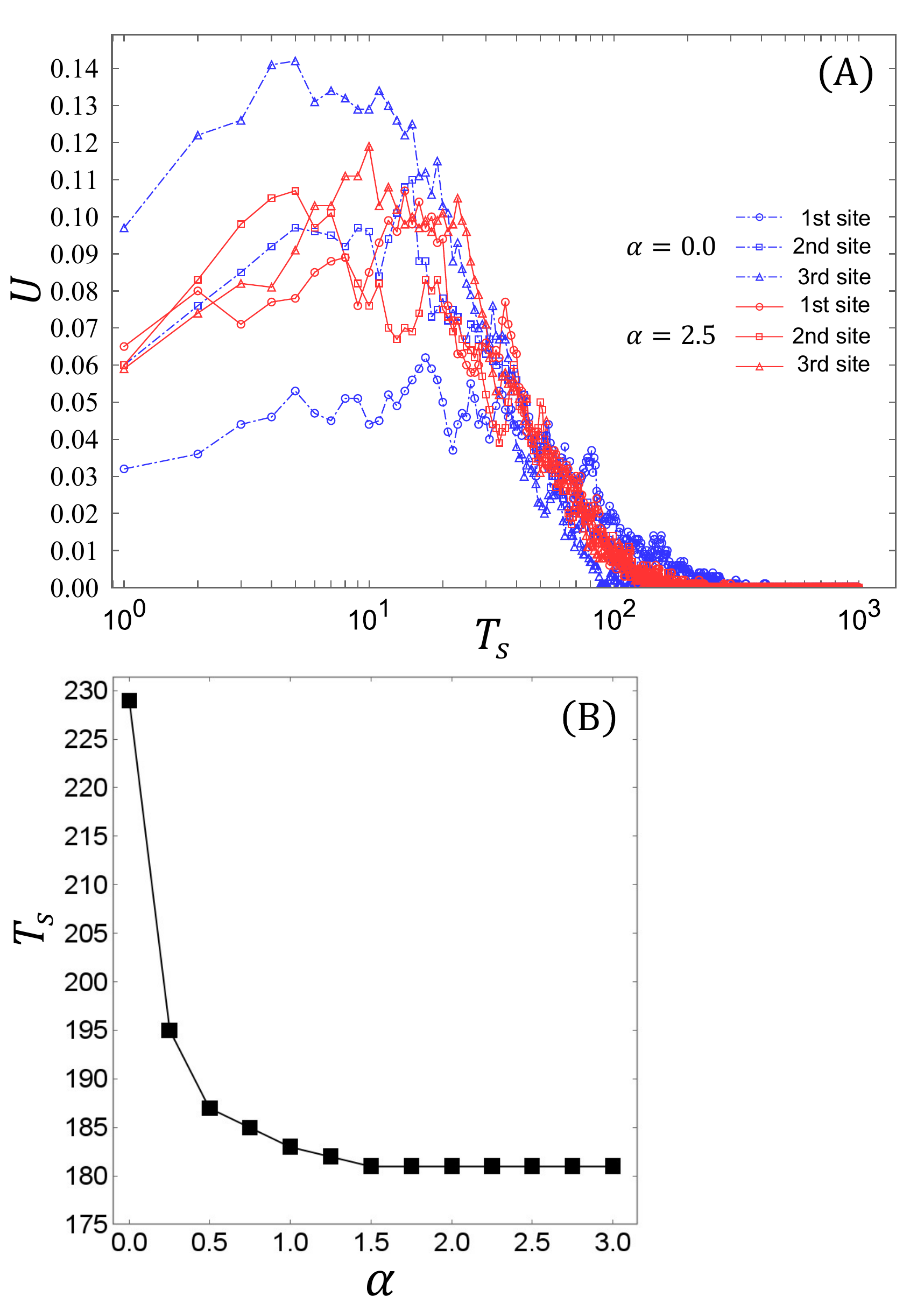}
\caption{(A) Change of usage $U$ of each node during the time taken for all agents to finish traveling to all the nodes, and (B) Dependence of the travel time for all agents visiting all the secondary nodes on the parameter $\alpha$.}
\label{fig:Figure_UChangeRate_TravelTime}
\end{figure}

As a preliminary test for Scenario-2, we investigated $T_s$, the number of time steps required for all agents to complete visiting all secondary nodes for the system with $M = 3$, similarly to Scenario-1. We set the parameters ($N_v$, $M$, $A$, $l$, $\alpha$) to (1,000, 3, 0.1, 2.0, 2.5). Figure~\ref{fig:Figure_UChangeRate_TravelTime}(A) shows the change in the usage $U$ of each secondary node during $T_{s}$. 
The blue chain lines indicate the results in the case of $\alpha=0$, where each agent stochastically chooses its destination by referring only to fixed preferences. By contrast, the red solid lines represent the results when $\alpha = 2.5$, where each agent stochastically chooses its destination by referring to the preferences declined by the congestion rate of each secondary node. 
The circular, square, and triangular symbols represent the first, second, and third nodes, respectively. 
Importantly, usage $U$ tends to change moderately and is distributed equally among the three nodes in the case of $\alpha = 2.5$ compared to the case of $\alpha = 0$. 
This result indicates that congestion avoidance by agents can mitigate the imbalance of usage among secondary nodes in Scenario-2, facilitate the effective use of secondary nodes and improve the travel efficiency of the network. 
Figure~\ref{fig:Figure_UChangeRate_TravelTime}(B) shows the dependence of $T_s$ on the parameter $\alpha$. We can confirm that $T_{s}$ drastically decreases soon after $\alpha$ becomes greater than zero, reaching a plateau at around $\alpha > 1.5$. 

In summary, it was confirmed from two preliminary tests for Scenario-1 and Scenario-2 that avoiding congestion can promote the equalization effect in both steady-state agent distribution and travel efficiency. Accordingly, in the following sections, we examine the equalization effect for general cases through simulations over a wider range of the number of secondary nodes, $M$, agents $N_{v}$, and the parameter $\alpha$. 

\begin{figure*}[t]
\vspace{-12.8cm}
\hspace{1.75cm}
\includegraphics[width=1.7\textwidth, clip, bb= 0 0 1571 1123]{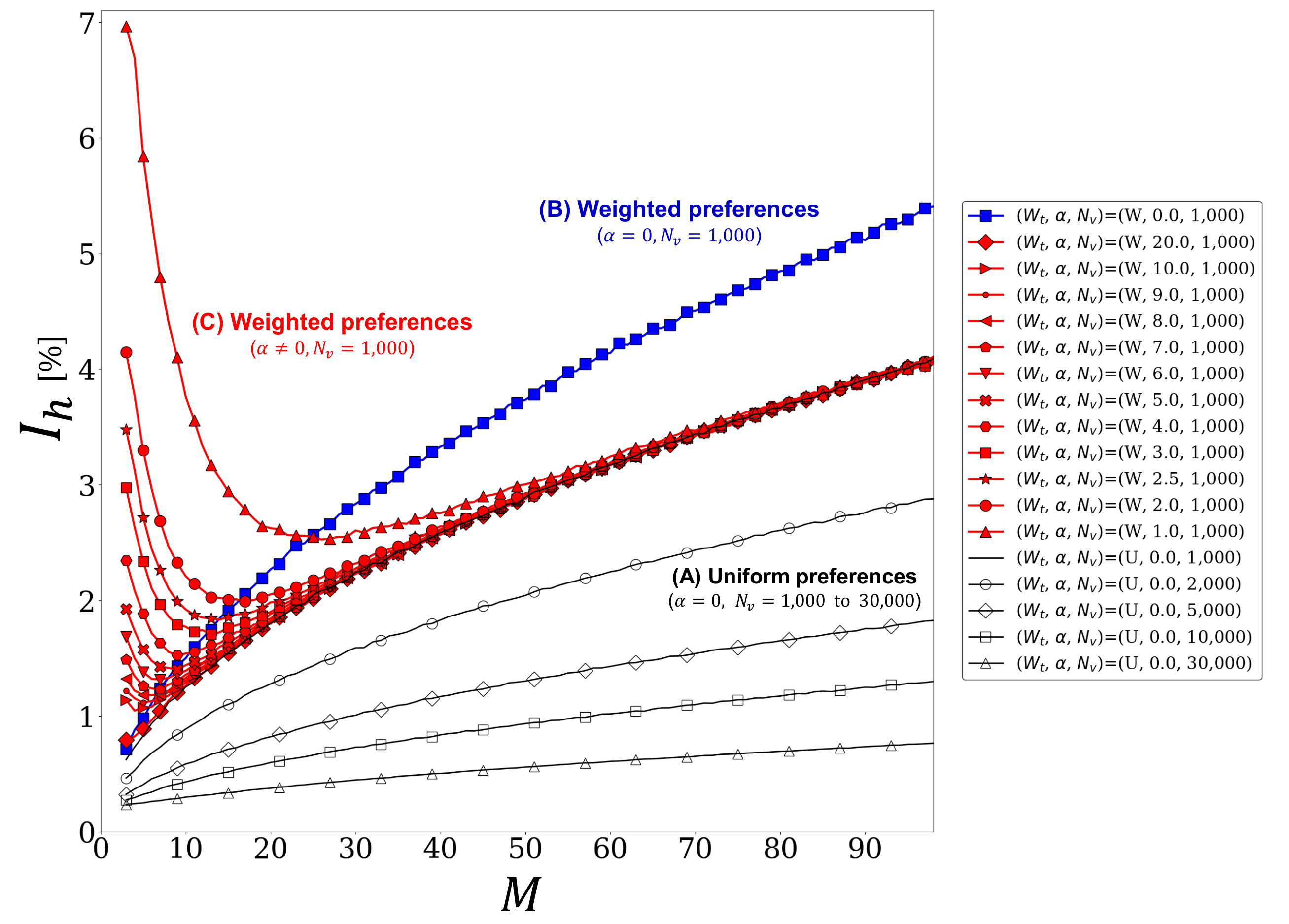}
\caption{Dependence of imbalance ratio $I_h$ on the parameter $M$ in different values of $N_v$ and $\alpha$ in Scenario-1}
\label{fig:Figure_Dependence_IhonM}
\end{figure*}
\section{Results}
\subsection{Simulation results for Scenario-1}
Figure~\ref{fig:Figure_Dependence_IhonM} shows the dependence of imbalance ratio $I_h$ on the parameter $M$ in different values of $N_v$ and $\alpha$ in Scenario-1. 
The dependence of $I_{h}$ on $M$ can be divided into three cases. 
The black line shows the case of uniform preferences with $\alpha = 0$, where we set weight $w_{i}$ to $1$ for all $i$.
The blue line shows the case of weighted preferences with $\alpha = 0$, where we set the weight $w_{i}$ to $i$ for the $i$th secondary node.  
The red lines show the cases of weighted preferences with $\alpha \ne 0$, where we set the weight $w_{i}$ to $i$ for the $i$th secondary node, as in case (B).
The results for the black, blue, and red colors are called (A), (B), and (C), respectively, in Fig.~\ref{fig:Figure_Dependence_IhonM}.
In the legend, ($W_t$, $\alpha$, $N_v$) represents the set of preference types, the parameter $\alpha$, and the number of agents $N_v$, where $W$ indicates weighted preferences and $U$ represents uniform preferences.
Specifically, in case (A), we measured the dependence of $I_{h}$ on parameter $M$ for different values of $N_{v}$ between $1,000$, $2,000$, $5,000$, $10,000$, and $30,000$ while keeping $\alpha$ constant at zero. It was observed that $I_{h}$ decreases as $N_{v}$ increases, and we obtain four different black curves. In case (B), we set the same parameters as (A) for $N_{v}=1,000$, except for the preferences mentioned above. In this case, $I_h$ was observed to be greater than in the case of (A) with $N_{v}=1,000$ in all areas of parameter $M$. In case (C), we set the same conditions as in case (B), except for $\alpha \ne 0$. Consequently, $I_h$ surges in a small area of $\alpha$ and increases as its minimum value $M$ increases. Thereafter, $I_h$ converges to case (A) when $N_{v}=1,000$, regardless of the degree of $\alpha$, and the strength of the surges in $I_h$ becomes larger as $\alpha$ decreases.
The reasons for the peculiar dependencies shown in Fig.~\ref{fig:Figure_Dependence_IhonM} from the viewpoint of the multivariate statistics are discussed in Section~\ref{sec:discuss}.

\begin{figure*}[t]
\vspace{-15.0cm}
\hspace{1.65cm}
\includegraphics[width=1.6\textwidth, clip, bb= 4 4 1248 1093]{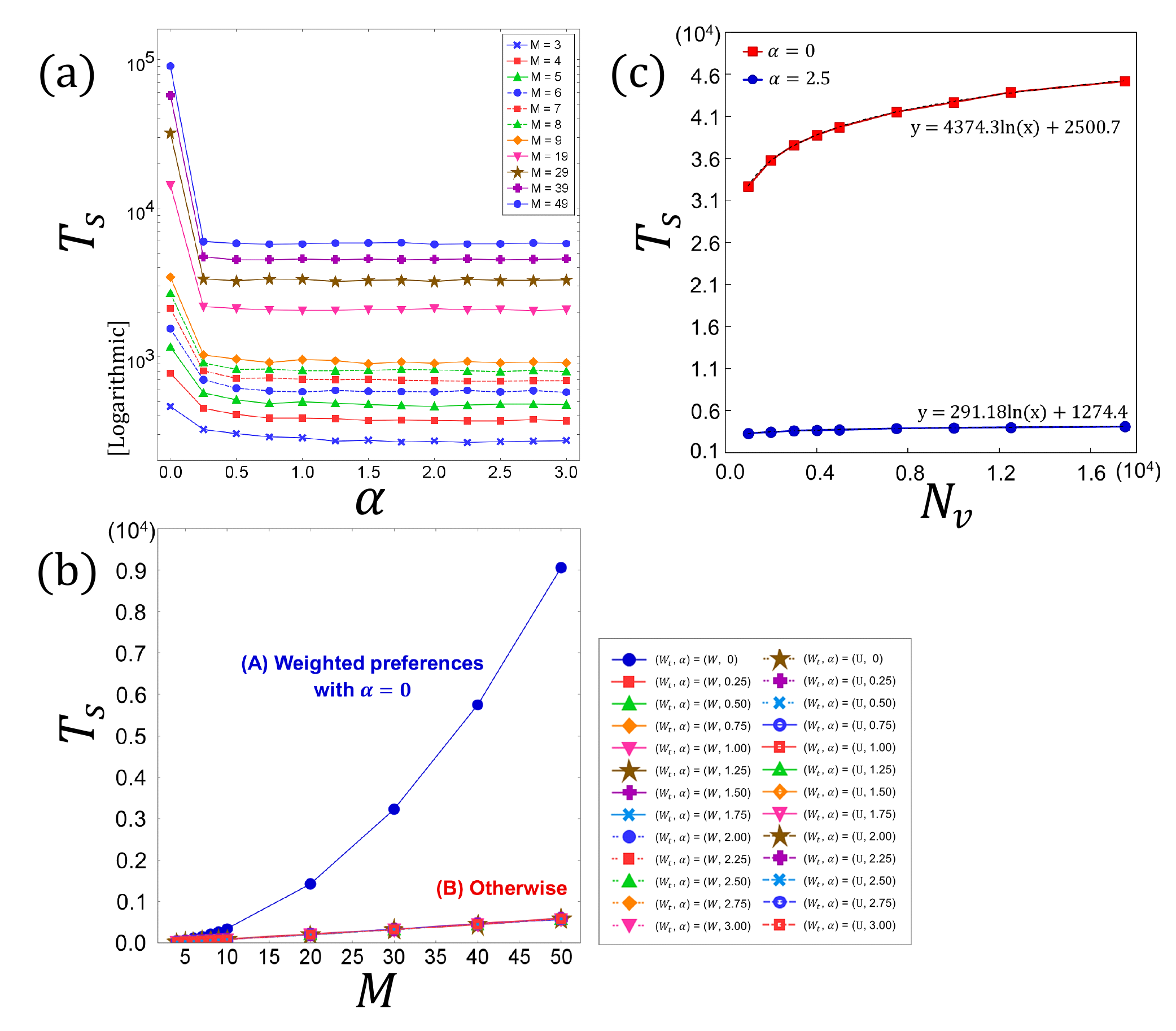}
\caption{Dependence of travel time steps $T_s$ on parameters $\alpha$, $M$, and $N_v$ in Scenario-2}
\label{fig:Figure_Scenario2}
\end{figure*}
\subsection{Simulation results for Scenario-2}
We measured the dependence of travel time steps $T_{s}$ on parameters $\alpha$, $M$, and $N_{v}$ in Scenario-2. First, we set $N_{v}$ to $1,000$ and then measured the dependence of $T_{s}$ on parameters $\alpha$, and $M$. Figure~\ref{fig:Figure_Scenario2}(a) shows the dependence of $T_{s}$ on parameter $\alpha$ for four different values of $M$ from $3$ to $49$ on the semilogarithmic scale in the case of weighted preferences, where we set the weight $w_{i}$ to $i$ for the $i$th secondary node. As observed in the preliminary tests with $M=3$, $T_{s}$ decreases soon after $\alpha$ becomes greater than zero and reaches a plateau in all cases of $M$. It was also observed that the difference in $T_s$ between the cases of $\alpha=0$ and $\alpha \ne 0$ increases as $M$ increases. Figure~\ref{fig:Figure_Scenario2}(b) shows the dependence of $T_{s}$ on the parameter $M$ in various $\alpha$ ranges between $0$ and $3.0$ on the linear scale in the cases of both weighted and uniform preferences. Notably, $T_{s}$ was observed to increase exponentially when the preferences were weighted with $\alpha=0$. By contrast, $T_s$ linearly increased to a similar degree when the preferences were otherwise weighted with $\alpha \ne 0$ or uniform.
An interesting point is that we obtain the same results as uniform preferences when $\alpha \ne 0$, which is almost irrelevant to the degree of parameter $\alpha$.
In addition, Fig.~\ref{fig:Figure_Scenario2}(c) shows the dependence of $T_{s}$ on the number of agents $N_{v}$ for $\alpha = 0$ and for $\alpha = 2.5$, which is a representative case of $\alpha \ne 0$. By fitting each case using a function of $y=a{\rm log(x)}+b$ where $a$ and $b$ are constants, it was found that $T_{s}$ increased more moderately when $\alpha \ne 0$ compared to when $\alpha = 0$; We confirmed from Fig.~\ref{fig:Figure_Scenario2}(c) that the stability of the network in the $N_{v}$ direction also increased because of the equalization effect. 

\section{Discussion}\label{sec:discuss}
\subsection{Multivariate statistics for agent distribution on a star-topology}
The imbalance ratio $I_{h}$ in Eq.~(\ref{eq:inhomo}), $S_{h, i}^{n}$ is a deterministic value because $S_{h, i}^{n}$ holds only constant parameters: preference $w_{i}$, control parameter $\alpha$, and number of secondary nodes $M$. By contrast, $S_{i}^n$ is a stochastic variable because it holds $N_{i}(t)$, as confirmed from Eq.~(\ref{eq:considerconges}) and Eq.~(\ref{eq:statevecresultcasea}); accordingly, $I_h$ can be represented as a function of the set of stochastic variables $S_{i}^{n}$ ($i=1,2,\cdots,M$) as follows: 
\begin{eqnarray}
I_h(S_1, S_2, \cdots, S_i,\cdots, S_M), \label{eq:multivalmap}
\end{eqnarray}
where $S_{i}$ is an abbreviation for $S_{i}^{n}$ and $I_h$ is a mapping from the set of ($S_{1}$, $S_{2}$, $\cdots$, $S_M$) from a mathematical perspective. According to the multivariate statistics, the propagation of uncertainty $\sigma_{I_h}$ of $I_h$ is described as follows:
\begin{eqnarray}
\sigma_{I_h} = \sqrt{D}, \label{eq:expsigma} 
\end{eqnarray}
\begin{eqnarray}
D = \begin{pmatrix}
\frac{\partial I_h}{\partial S_1} \\
\frac{\partial I_h}{\partial S_2} \\
\vdots \\
\frac{\partial I_h}{\partial S_M} \\
\end{pmatrix}^t
\begin{pmatrix}
{\sigma_{1}}^2& {\sigma_{12}} & {\sigma_{13}} & \cdots & {\sigma_{1M}} \\
{\sigma_{21}} & {\sigma_2}^2  & {\sigma_{23}} & \cdots & {\sigma_{2M}} \\
{\sigma_{31}} & {\sigma_{32}} & {\sigma_3}^2  & \cdots & {\sigma_{3M}} \\
\vdots        &    \vdots     &      \vdots   & \ddots &    \vdots     \\
{\sigma_{M1}} & {\sigma_{M2}} & {\sigma_{M3}} & \cdots & {\sigma_M}^2 \\
\end{pmatrix}
\begin{pmatrix}
\frac{\partial I_h}{\partial S_1} \\
\frac{\partial I_h}{\partial S_2} \\
\vdots \\
\frac{\partial I_h}{\partial S_M} \\
\end{pmatrix},
\label{eq:valcovmat}
\end{eqnarray} 
where ${\partial I_h}/{\partial S_i}$ represents the partial derivative of $I_h$ with respect to variable $S_i$. The center bracket represents the variance-covariance matrix of the system (hereafter referred to as $\vec{E}$). ${\sigma_{i}}^2$ represents the variance of $I_h$ with respect to variable $S_{i}$, and $\sigma_{ij}$ indicates the covariance of $I_h$ between variables $S_{i}$ and $S_{j}$. Equation~(\ref{eq:expsigma}) can be expressed in scalar form as
\begin{eqnarray}
\sigma_{I_h} = \sqrt{\sum_{i=1}^{M} \biggl(\frac{\partial I_h}{\partial S_i}\biggr)^{2}\sigma_{i}^2 
	         + \sum_{i \ne j}^{M} \biggl(\frac{\partial I_h}{\partial S_i}\biggr) \biggl(\frac{\partial I_h}{\partial S_j}\biggr)~\sigma_{ij} }, \label{eq:expsigmascalar} 
\end{eqnarray}
The first and second terms inside the square root of Eq.~(\ref{eq:expsigmascalar}) are the contributions of the diagonal and non-diagonal components of the variance-covariance matrix $\vec{E}$, respectively. In matrix $\vec{E}$, the covariance $\sigma_{ij}$ becomes zero when the variables $N_{i}$ and $N_{j}$ are uncorrelated, and the second term inside the square root of Eq.~(\ref{eq:expsigmascalar}) vanishes if all the variables are uncorrelated. 

According to the theory of errors, the measured value $I_{h}$ can be decomposed into the mean value $\langle I_h \rangle$ and the uncertainty $\sigma_{I_h}$ as $I_{h}$ = $\langle I_h \rangle + \sigma_{I_h}$. When $\alpha=0$, $S_{i}$ and $S_{h, i}^n$ become equal since the second terms in $r_i$ in Eq.~(\ref{eq:considerconges}) and $\langle r_{i} \rangle$ in Eq.~(\ref{eq:statevecapprox}) vanishe; from the definition of $I_h$ in Eq.~(\ref{eq:inhomo}), the mean $\langle I_{h} \rangle$ can be estimated as zero. Meanwhile, when $\alpha \ne 0$, the number of agents in each secondary node is equalized after reaching a stationary state owing to the equalization effect. Accordingly, we can assume that the difference between $S_{i}^{n}$ and $S_{h, i}^n$ becomes sufficiently small to be negligible, and the relationship of $S_{i}^{n} \approx S_{h, i}^n$ can be established for a sufficiently large $\alpha$. In other words, $\langle I_h \rangle$ can be approximately zero or a small number $\epsilon$. In summary, we assume the following relationship:
\begin{eqnarray}
I_h = 
\begin{cases}
\sigma_{I_h}& (\alpha = 0)\\
\sigma_{I_h} + \epsilon& (\alpha \ne 0)
\end{cases} \label{eq:decompIh}
\end{eqnarray}

When $\alpha=0$, each agent chooses its direction only by referencing fixed preferences. In this case, all resulting statistics of ($N_1$, $N_2$, $\cdots$, $N_M$) in the secondary nodes are uncorrelated. Because of the linear transformation relationship, all stochastic variables ($S_1$, $S_2$, $\cdots$, $S_M$) are also uncorrelated. Accordingly, the second term in the square root of Eq.~(\ref{eq:expsigmascalar}) vanishes,  the covariances in the non-diagonal components of $\vec{E}$ are zero; we obtain the value of parameter $\sigma_{I_h}$ only from the first term in Eq.~(\ref{eq:expsigmascalar}). Because we can calculate ${\partial I_{h}}/{\partial S_i}$ by differentiating Eq.~(\ref{eq:inhomo}) with respect to $S_{i}$, the remaining parameter that must be estimated is the deviation $\sigma_{i}$. 
\HLT{By contrast, when $\alpha \ne 0$, we need to estimate $\sigma_{ij}$ in addition to $\sigma_{i}$ because of the emergence of the correlations among secondary nodes.}

\HLT{We refer to ~\ref{sec:appderivation} for the details of the mathematical derivations of $\sigma_{i}$ and $\sigma_{ij}$}. 
\HLT{The resulting expressions of the deviation $\sigma_{i}$ in the respective cases are as follows:}
\begin{eqnarray}
\sigma_{i}^{A} &\approx& \sigma_{0}\sqrt{N_v M {S_{h,i}^n} (1- {S_{h,i}^n})}, \label{eq:estimsigmaA-body} \\
\sigma_{i}^{B,~C} &\approx& \mu M (S_{h,i}^{n}-b) \sqrt{N_v M {S_{h, i}^{n}} (1- {S_{h, i}^{n}})}. \label{eq:estimsigmaB-body} 
\end{eqnarray}
\HLT{In addition, the covariance $\sigma_{ij}$ for case (C) is given as follows:}
\begin{eqnarray}
\sigma_{ij} \approx C_d\biggl\{ \bigl(\bar{c}~S_{h,i}^{n} S_{h,j}^{n}\bigr)^2 - \frac{1}{(l+1)^2M^2}\biggr\}. \label{eq:estimsigmaC-body}
\end{eqnarray}
Our \HLT{theoretical} model has scale parameters of $\sigma_{0}$ for case (A), $\mu$ for case (B), and ($\bar{c}$, $C_d$, $\mu$) for case (C). 
We determined these scale parameters by fitting the experimental data \HL{because the scale of the system state is indeterminate, as represented by parameter $c$ in Eq.~(\ref{eq:statevecresultcasea})}.  
\HL{Specifically}, in case (A), we first determined $\sigma_{0}$ by fitting the case of $N_v=1,000$ using least squares and used the same value when plotting model value of $I_{h}$ in other cases of $N_v=2,000$, $N_v=5,000$, $N_v=10,000$, and $N_v=30,000$. In case (B), we determined $\mu$ in a manner similar to that in case (A). In case (C), we searched for the optimal condition of ($\bar{c}$, $C_d$, $\mu$) that reproduces the measurements shown in Fig.~\ref{fig:Figure_Dependence_IhonM}. In addition, we investigated the dependence of the phenomenological scale parameter $C_d$ on parameter $\alpha$. 

\begin{figure*}[t]
\vspace{-15.0cm}
\hspace{+1.0cm}
\includegraphics[width=1.8\textwidth, clip, bb= 0 0 2029 1631]{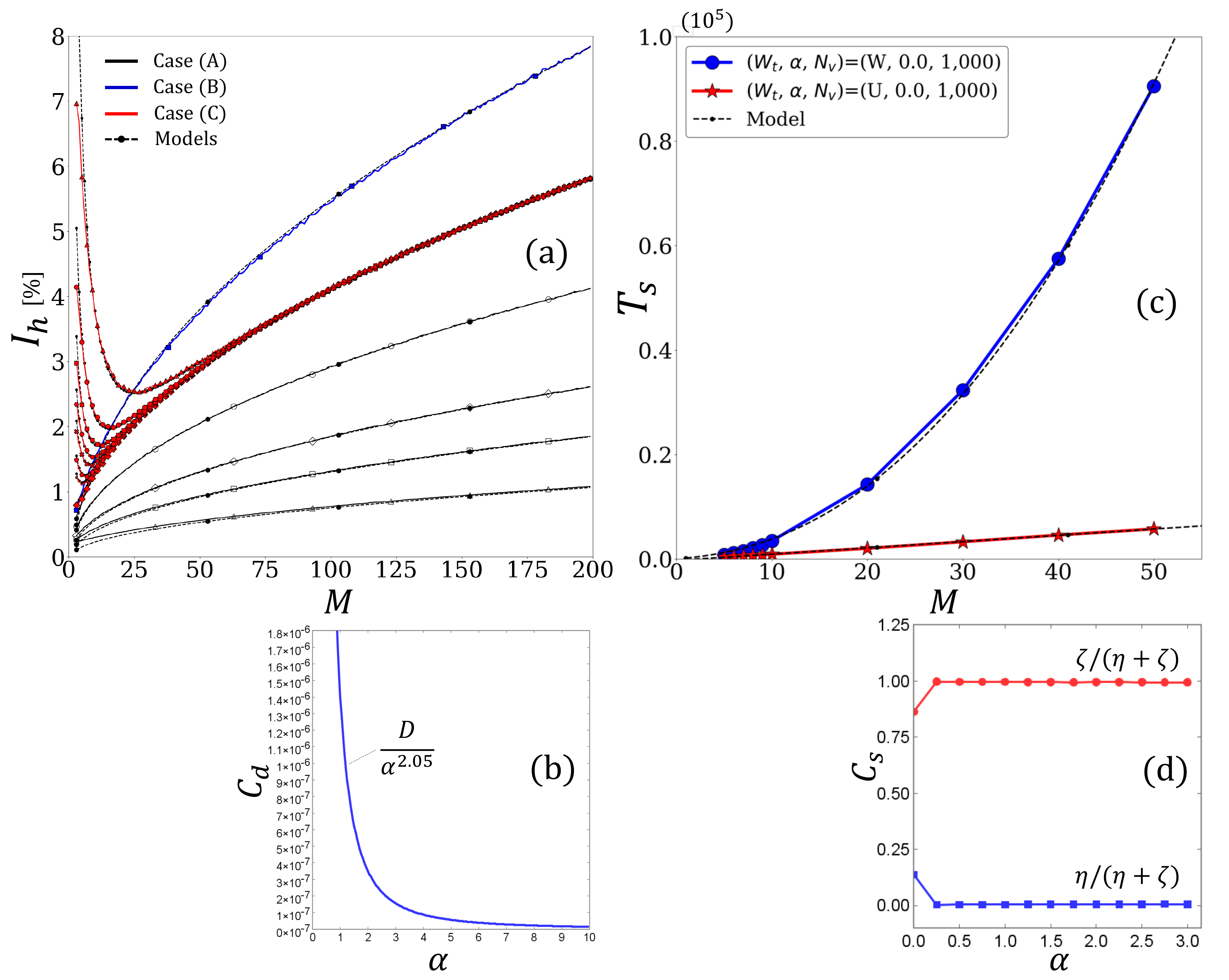}
\caption{(a) Comparison of our models with the simulation results in cases of (A), (B), and (C) in Fig.~\ref{fig:Figure_Dependence_IhonM} in Scenario-1, (b) dependence of parameter $C_{d}$ on parameter $\alpha$, (c) comparison of our models with the simulation results in two cases of uniform preferences and the weighted preferences with $\alpha=0$ in Scenario-2, and (d) dependence of parameter $C_s$ on parameter $\alpha$}
\label{fig:Figure_CompareModelSimScen2}
\end{figure*}

Figure~\ref{fig:Figure_CompareModelSimScen2}(a) shows the comparisons of our \HL{theoretical} models from Eq.~(\ref{eq:estimsigmaA-body}) to Eq.~(\ref{eq:estimsigmaC-body}) with the measurements shown in Fig.~\ref{fig:Figure_Dependence_IhonM}. Each solid line corresponds to the result with the same color and symbol in Fig.~\ref{fig:Figure_Dependence_IhonM}. The \HL{theoretical} models show good agreement with measurements in all three cases from (A) to (C). The resulting $\sigma_{0}$ obtained by fitting the measurement in the case of (A) with $N_{v} = 1,000$ was $1.316 \times 10^{2}$ in this test. The other plots for various $N_{v}$ in case (A) were obtained using the determined $\sigma_{0}$. As a result of fittings, the optimal values of $\mu$ for case (B) and $(\bar{c}, \mu)$ for case (C) were obtained as $3.072\times 10^{-4}$ and ($1.0 \times 10^{3}$, $2.258 \times 10^{-4}$), respectively. Note that the parameter $\epsilon$ in Eq.~(\ref{eq:decompIh}) was confirmed to be zero. 

The model \HL{in} case (B) calculates only the diagonal components of the variance-covariance matrix $\vec{E}$. \HL{On the other hand}, the model \HL{in} case (C) calculates the non-diagonal components of $\vec{E}$ as well as the diagonal components \HL{using a generic statistical relationship:~$\sigma_{ij} = \langle S_{i} S_{j} \rangle - \eta_{i}\eta_{j}$} since \HL{it is expected that} the statistics in different secondary nodes correlate from each other \HL{as a result of} agents mutually referring to the congestion rate of the secondary nodes. \HL{In this respect}, we \HL{can confirm from Fig.~\ref{fig:Figure_CompareModelSimScen2}(a)} that the \HL{contributions from the} non-diagonal components of $\vec{E}$ \HL{reproduce} the characteristic surges of $I_{h}$ in small areas of the parameter $M$ \HL{and converge to the case of uniform preferences}. \HLL{This suggests that mutually referring to the congestion information primarily causes the surges of $I_h$, that is, the deterioration of uniformity. In other words, referencing congestion information can make the uniformity worse rather than better if the degree of reference to the congestion information is insufficient in small systems.} 

\HLL{As mentioned in the Introduction section, the network becomes endogenous when the system provides agents with congestion information to balance agents among nodes because the resulting agent distribution feeds back into the input congestion rates; hence, our findings can be helpful when controlling such endogenous traffic networks that provide agents with congestion information as traffic information in real-world cases.}
In addition, Fig.~\ref{fig:Figure_CompareModelSimScen2}(b) shows a plot of the values of $C_{d}$ for different values of $\alpha$ in case (C), where $D=1.3 \times 10^{-6}$. It was found that parameter $C_{d}$ is approximately proportional to the reciprocal of the square of parameter $\alpha$. We can confirm the following from Fig.~\ref{fig:Figure_CompareModelSimScen2}(b). When agents avoid congestion linearly to the congestion rate with the scale of $\alpha$, the uniformity of the network varies approximately depending on the inverse of the square of the parameter $\alpha$. This finding \HL{also} serves as a guide for applying our results to engineering. 

Consequently, our theoretical model accurately describes the mechanism of the target system. Importantly, our analysis \HL{corroborated} that the balance between the equalization of network usage by avoiding congestion and the amplification of covariance caused by a mutual reference to congestion information determines the overall uniformity of a network with star topology.

\subsection{ Traveling time in Scenario-2}
The most straightforward way to reproduce the travel time steps $T_{s}$ in Scenario-2 is to assume that the times for moving from the primary to the secondary node and returning from the secondary node are proportional to the reciprocals of the hopping and return probabilities. For example, when uniform preferences with $\alpha=0$, the hopping probability is given by $1/M$ for each secondary node, and the returning probability is given by $lA$, as explained in Section~\ref{sec:targetsysmodel}. Hence, the travel time step required to visit a cell, $t_s$, can be described as $t_{s} = \eta M^{-1} + \zeta {lA}^{-1}$, where $\eta$ and $\zeta$ are the coefficients. By summing $t_s$ from $1$ to $M$, we can obtain the expression of $T_{s}$ as follows:
\begin{eqnarray}
T_s = \eta + \zeta \frac{M}{lA} \label{eq:traveltimeuni}
\end{eqnarray}
In the case of weighted preferences, we can calculate the travel time steps $T_{s}$ similarly for uniform preferences after replacing $1/M$ with the inverse of the probability $S_{h, i}^n$. Specifically, 
\begin{eqnarray}
T_s = \eta \sum_{i=1}^{M} \frac{1}{S_{h,i}^{n}} + \zeta \frac{M}{lA}, \label{eq:traveltimeweighted}
\end{eqnarray}
When $\alpha=0$, Eq.~(\ref{eq:traveltimeweighted}) can be further broken down as 
\begin{eqnarray}
T_s &=& \eta \frac{(l+1)M(M+1)}{2} \sum_{i=1}^{M} \frac{1}{i} + \zeta \frac{M}{lA} \nonumber \\
	&\approx& \eta \frac{(l+1)M(M+1)}{2} \biggl({\rm ln} M + \gamma + \frac{1}{2M} \biggr) + \zeta \frac{M}{lA}, \label{eq:traveltimeweightedsimple} 
\end{eqnarray}
where we use the mathematical relationship of $\sum_{i=1}^{k} 1/i \approx {\rm ln} k + \gamma + 1/2k$; $\gamma$ represents the Euler--Mascheroni constant, which nearly equals $0.577215664$~\cite{CHEN2012391}. 

In Fig.~\ref{fig:Figure_CompareModelSimScen2}(c), the solid red line with the star symbol and the solid blue line with the circle symbol, respectively, indicate the measurements of uniform preferences and weighted preferences with $\alpha=0$. The dashed lines represent the fitting lines obtained using Eq.~(\ref{eq:traveltimeweighted}) and Eq.~(\ref{eq:traveltimeweightedsimple}). We confirmed that our model describes the variation in the traveling time step in Scenario-2. Fig.~\ref{fig:Figure_CompareModelSimScen2}(d) shows the dependence of the ratio of $\eta$ or $\zeta$ on the sum of the parameters $\alpha$ in the case of the weighted preferences obtained by fitting \HL{the measurements by our model in Eq.~(\ref{eq:traveltimeweightedsimple})} for all cases of $\alpha \ne 0$, where $C_{d}$ represents $\eta/(\eta+\zeta)$ or $\zeta/(\eta+\zeta)$. It was confirmed that the relative ratio of $\zeta$, which is the coefficient of the return time, increases, and the relative ratio of $\eta$, which is the coefficient of the hopping time from primary to secondary, decreases; \HL{the time required for the outward trip was confirmed to decrease as a result of agents avoiding congestion due to the increase in parameter $\alpha$}. 

\begin{figure*}[t]
\vspace{-10.0cm}
\hspace{+1.0cm}
\includegraphics[width=1.83\textwidth, clip, bb= 10 10 2002 1127]{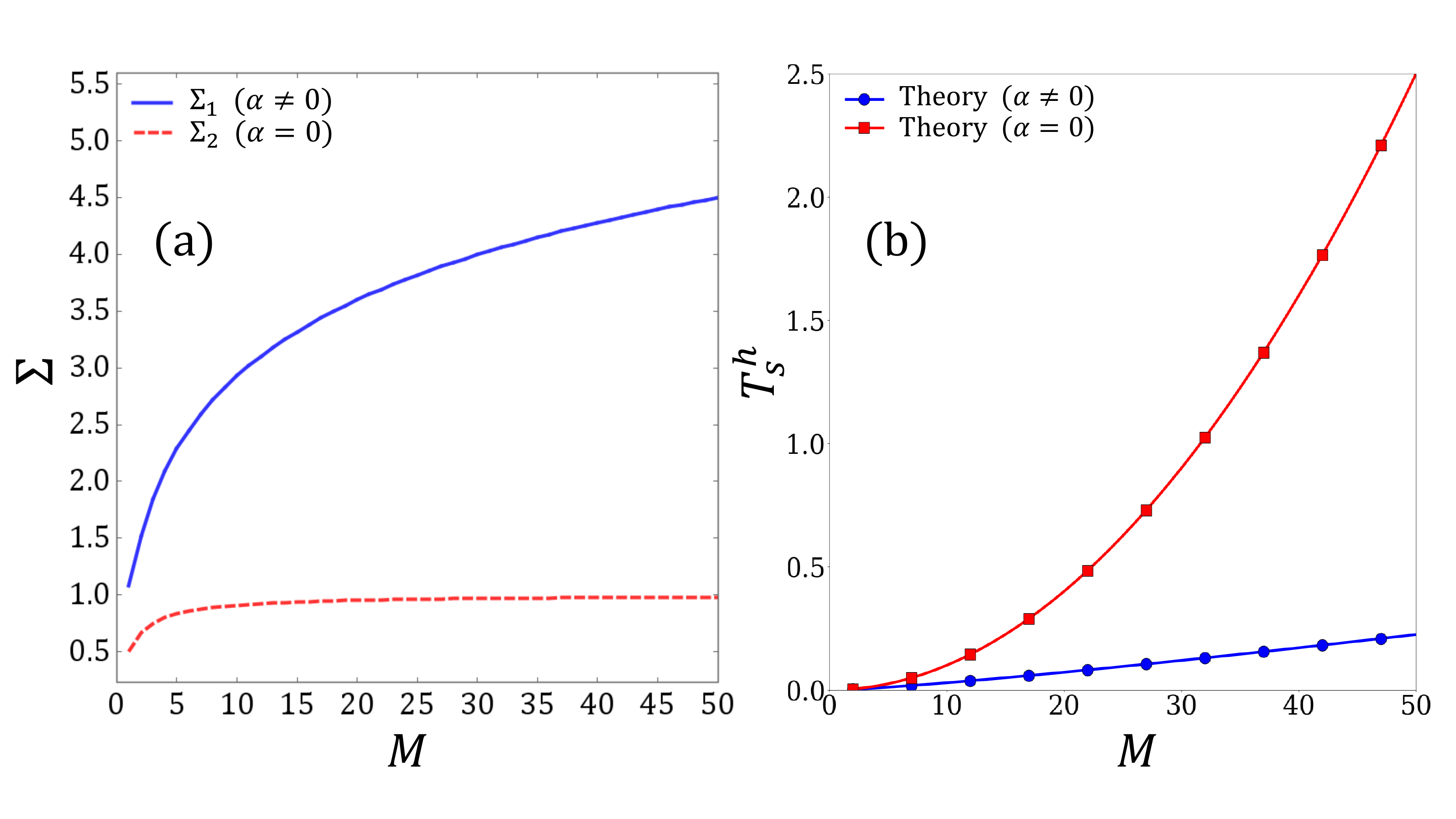}
\caption{
(a) Dependence of $\Sigma_1$ and $\Sigma_2$ on parameter $M$, and (b) dependence of the total time steps $T_s^h$ required for hopping from the primary to secondary nodes on parameter $M$ when weighted preferences in respective cases of $\alpha = 0$ and $\alpha \ne 0$
}
\label{fig:Figure_CompareModelLinearExp}
\end{figure*}

\HLL{In addition, we can show the exponentiality and linearity of time steps $T_s$ in the respective cases of weighted preferences with $\alpha=0$ and $\alpha\ne0$ from a different angle as follows: As mentioned in Section~\ref{sec:targetsysmodel}, the total probability is normalized for all the secondary nodes, and agents lose the probabilities of choosing the already-visited secondary nodes; if an already-visited node is selected in a trial, the system discards the trial. Therefore, the probability of hopping to any of the secondary nodes at the time of $k$th visiting, $P_{u}$, is given as a complement of the sum of the probabilities of hopping to one of the secondary nodes that have been already visited before the $k$th visiting, $P_{a}$, that is, $P_{u}+P_{a}=1$, where the subscripts $u$ and $a$ indicate the ``unvisited'' and ``already-visited'' nodes, respectively. When an agent avoids congestion owing to $\alpha \ne 0$, the probability of hopping to one of the secondary nodes becomes equalized and can be approximately expressed as $1/M$; $P_a$ at the time of the $k$th visit $(k=1,2\cdots, M)$ is given as $k/M$, and $P_u$ is expressed as $(1-k/M)$. Here, we assume that the average time step required to hop from the primary node to one of the secondary nodes at the $k$th visit, $t_s^{k}$, is proportional to the inverse of the value of $P_u$ at the $k$th visit; $t_s^{k}$ can be expressed as $\eta (1-k/M)^{-1}$. Accordingly, the total time steps required for hopping from the primary node to the secondary nodes, $T_{s}^{h}$, can be obtained by summing up $t_s^{k}$ for $k$; A simple calculation leads to:}
\begin{eqnarray}
\HLL{T_{s}^{h} = \eta M \sum_{k=1}^{M}\frac{1}{k}. \label{eq:micromodeluni} }
\end{eqnarray}
\HLL{Here, the summation in Eq.~(\ref{eq:micromodeluni}) can be approximated as $\sum_{k=1}^{M} 1/k \approx {\rm ln} M + \gamma + 1/2M$, similar to Eq.~(\ref{eq:traveltimeweightedsimple}). We refer to the summation in Eq.~(\ref{eq:micromodeluni}) as $\Sigma_{1}$. As we can observe from Euler's approximation, $\Sigma_{1}$ shows a moderate dependence on the parameter $M$, as shown by the blue line in Fig.~\ref{fig:Figure_CompareModelLinearExp}(a). Consequently, the component $\eta M$ of Eq.~(\ref{eq:micromodeluni}) is confirmed to be dominant in the $M$-dependence of $T_s^h$, as indicated by the blue line with the circle symbol in Fig.~\ref{fig:Figure_CompareModelLinearExp}(b). Accordingly, the linearity of $T_{s}$ is confirmed for weighted preferences with $\alpha \ne 0$.}

\HLL{Meanwhile, when $\alpha = 0$, agents move to one of the secondary nodes only according to the fixed preference $i$ for the $i$th secondary node. Because the secondary node with a larger weight is preferentially visited, $P_a$ at the time of the $k$th visit $(k=1,2\cdots, M)$ can be expressed as $\sum_{j=1}^{k}(M-j+1)/W$ as a typical case, where $W$ is the sum of the preference $i$ from $1$ to $M$, which is $M(M+1)/2$. In this case, $P_u$ is expressed as $1-\sum_{j=1}^{k}(M-j+1)/W$. We assume that the average time steps $t_{s}^k$ is proportional to the inverse of the value of $P_u$ at the $k$th visit, similar to Eq.~(\ref{eq:micromodeluni}). Then, $T_s^h$ is calculated by summing $t_s^k$ for $k$; the resulting expression of $T_{s}^{h}$ is obtained as follows:} 
\begin{eqnarray}	
\HLL{T_{s}^{h} = \eta M(M+1)\sum_{i=0}^{M-1}\frac{1}{M(M+1)-i(2M-i+1)}. \label{eq:micromodelweighted}} 
\end{eqnarray}	
\HLL{
Here, we refer to the summation of Eq.~(\ref{eq:micromodelweighted}) as $\Sigma_{2}$: The dashed red line in Fig.~\ref{fig:Figure_CompareModelLinearExp}(a) shows the dependence of $\Sigma_{2}$ on the parameter $M$, which becomes sluggish at approximately $M=25$, and then converges to one. Consequently, component $\eta M(M+1)$ in Eq.~(\ref{eq:micromodelweighted}) is confirmed to be dominant in the $M$-dependence of $T_{s}^h$, as indicated by the red line with the square symbol in Fig.~\ref{fig:Figure_CompareModelLinearExp}(b). Accordingly, the exponentiality of $T_{s}$ is confirmed for weighted preferences, with $\alpha = 0$.
Notably, we describe all the cases observed in Fig.~\ref{fig:Figure_Scenario2}(b) using the theoretical models represented by Eq.(\ref{eq:traveltimeuni}) into Eq.(\ref{eq:micromodelweighted}): Uniform preferences with $\alpha=0$, weighted preferences with $\alpha=0$, and weighted preferences with $\alpha\ne0$.}

\HLLL{
A star topology is often used as a conceptual decision-making model by individuals with multiple choices. The relationship between $P_u$ and $P_a$ is similar to that between the fatigue from visiting congested nodes and the motivation to leave the primary node; $P_a$ increases by approximately $1/M$ every time returning to the primary node when the preferences were weighted with $\alpha\ne0$ or uniform. By contrast, $P_a$ is $\sum_{j=1}^{k} (M-j + 1)/W$ at the time of the $k$th visit when weighted preferences with $\alpha=0$ because every agent tends to visit the nodes with higher preferences. Namely, the nodes with higher preferences get more congested. If we assume that the fatigue from visiting a node is proportional to the degree of congestion in the node, $P_a$ represents accumulated fatigue due to visiting congested nodes. Our results suggest that fatigue moderately increases when $\alpha \ne 0$ because agents can avoid congestion; however, it drastically builds up when $\alpha = 0$ because they face congestion. In reality, agents are not necessarily forced to visit secondaries, and thus turnover rate of secondary nodes can decline when they are tired; Our results can explain the importance of presenting congestion information from a mental health perspective.}
Consequently, we succeed in \HL{elucidating} the effect of congestion avoidance on the uniformity and travel efficiency of a traffic network with star topology. 

\section{Conclusion}
Fundamental research on network topologies has been widely conducted in many scientific areas. This study presented a multivariate statistical analysis of the effect of congestion-avoiding behavior owing to congestion information provision on optimizing traffic in star topologies. We investigated the dynamics of a stochastic transportation network in which each agent at the primary node stochastically chooses one of the secondary nodes by referring to fixed preferences, which are reduced by the congestion rate of each secondary node. We examined the following two scenarios: each agent can repeatedly access the same secondary node, or each agent can access each secondary node only once. We refer to the former scenario as Scenario-1 and the latter as Scenario-2.
We examined the uniformity of agent distribution in the stationary state in Scenario-1, and we measured the travel time for all agents visiting all the nodes in Scenario-2. 
\HL{The findings of this study are summarized as follows:}

\HL{In the case that agents repeatedly visit central and other nodes in Scenario-1, the uniformity of agent distribution was found to show three types of nonlinear dependences on the increase of nodes. We clarified that multivariate statistics describe these characteristic dependences well}. Our theoretical analysis \HL{corroborates} that the balance between the equalization of network usage by avoiding congestion and the amplification of covariance caused by mutual reference to congestion information determines the overall uniformity of the network with star topology. 
\HLL{This further suggests the following: Referencing congestion information can make the uniformity of networks worse rather than better if the degree of reference to the congestion information is insufficient in small systems; this finding can be helpful when controlling the endogenous traffic networks that provide agents with congestion information as traffic information in real-world cases.}
\HL{In Scenario-2,} we discovered that congestion-avoiding behavior linearizes the travel time, which exponentially increases with the number of nodes, notwithstanding the degree of reference to congestion information. 
\HLL{Our theoretical models clearly explain the linearity and exponentiality in the respective cases.}
Consequently, we successfully described the optimization effect of congestion-avoiding behavior on the collective dynamics of agents in star topologies. These fundamental perspectives are helpful in several areas of network science.

\section*{Acknowledgment}
This work was supported by the JST-Mirai Program Grant Number JPMJMI20D1, Japan, as well as JSPS KAKENHI Grant Numbers JP21H01570 and  21H01352.
The author would like to thank Editage (www.editage.jp) for the English language editing service. 
We express special thanks to the administrative staff in Nishinari laboratory and RCAST.
{\bf Author contributions}: S.T., D.Y., and K.N. designed the study; S.T. performed modeling and numerical experiments and analyzed the data; D.Y. and K.N. provided funding acquisition, project administration, and resources; and S.T drafted the manuscript. All authors contributed to the feedback and editing of the manuscript. {\bf Competing interests}: The authors declare no competing interests.



\appendix
\section{Details of the derivations of $\sigma_{i}$ and $\sigma_{ij}$} \label{sec:appderivation}
Each agent selects the $i$th node from among the $M$ secondary nodes with probability $r_{i}$ and avoids the $i$th node with probability $1 - r_{i}$ at each time step, as \HL{mentioned} in Section~\ref{sec:targetsysmodel}. Accordingly, the system follows a Bernoulli trial, where the deviation is represented by $\sigma = \sqrt{Np(1 - p)}$. $N$ indicates the number of statistics and $r$ represents the probability. 
In case (A), we can replace $p$ with $S_{h, i}^n$ because it represents the probability of finding an agent in the $i$th secondary node in a stationary state. In addition, the total number of statistics is \HLT{proportional to} $N_{v} M$, which is the number of secondary nodes multiplied by \HLT{the total number of} agents in the target system. \HLT{We introduce} a constant parameter $\sigma_{0}$. 
Consequently, an approximation for the uncertainty $\sigma_{i}$ is obtained as follows: 
\begin{eqnarray}
\sigma_{i} \approx \sigma_{0}\sqrt{N_v M {S_{h,i}^n} (1- {S_{h,i}^n})}, \label{eq:estimsigmaA}
\end{eqnarray}

In case (B), it is necessary to modify Eq.~(\ref{eq:estimsigmaA}) owing to weighted preferences. The serial indices are set to the secondary nodes as linearly weighted preferences; the $i$th node has weight $i/W_{s}$, where $W_{s}$ is the sum of the numbers from $1$ to $M$, which is $M(M +1)/2$. When two variables $X$ and $Y$ have a linear relationship $Y = aX + b$, their variances $\sigma_{Y}^{2}$ and $\sigma_{X}^{2}$ satisfy $\sigma_{Y}^2 = a^{2}\sigma_{X}^2$. Therefore, the variance of the $i$th secondary node for weighted preferences differs from that for uniform preferences. We estimate the variance in the $i$th secondary node as follows: We consider a linear transformation of $S_{h, i}^{n}$ from the state vector $\overline{S_{h, i}^n}$ as $a = (S_{h,i}^{n}-b) / \overline{S_{h, i}^{n}}$, where $\overline{S_{h, i}^{n}}$ is a state vector with uniform preferences of $\alpha=0$ obtained by setting $w_i$ to 1 for all $i$, which is expressed as $\overline{S_{h, i}^n} = 1/(l+1)M$. We \HL{modify} Eq.~(\ref{eq:estimsigmaA}) to be proportional to parameter $a$ as follows:
\begin{eqnarray}
\sigma_{i} \approx \mu M (S_{h,i}^{n}-b) \sqrt{N_v M {S_{h, i}^{n}} (1- {S_{h, i}^{n}})}, \label{eq:estimsigmaB}
\end{eqnarray}
where we introduce a constant factor $\mu$ in addition to the parameter $b$ for a \HL{simple} expression. 

In case (C), it is necessary to calculate the covariance between the secondary nodes. In this case, agents choose their destinations by referring to the congestion rates of all secondary nodes, which suggests that the statistics in a secondary node depend on the congestion status of the other secondary nodes. In other words, the statistics of the different secondary nodes correlate with each other. Accordingly, it is necessary to calculate both the diagonal and non-diagonal components of the variance-covariance matrix $\vec{E}$. We estimate the values of the covariance components of $\vec{E}$ as follows: First, we recall that there is a general relationship between the two correlating stochastic variables $S_i$ and $S_j$ as $\sigma_{ij}= \langle S_{i} S_{j} \rangle - \eta_{i}\eta_{j}$, where $\sigma_{ij}$ is the covariance between $S_i$ and $S_j$, $\eta_{i}$ and $\eta_{j}$ represent the means of $S_{i}$ and $S_{j}$, and $\langle S_{i} S_{j} \rangle$ is the mean of the multiples of $S_{i}$ and $S_{j}$. We evaluate $\eta_{x}~(x=i,j)$ by its arithmetic mean, which can be obtained by summing up $S_{h,i}^n$ for $i$ from $1$ to $M$ and dividing it by $M$; the resulting $\eta_{x}$ is $1/(l+1)M$, which yields the relationship $\sigma_{ij} = \langle S_{i} S_{j} \rangle - \{(l+1)M\}^{-2}$. 
Subsequently, because $\langle S_{i} S_{j} \rangle$ is the expected value of $S_{i}S_{j}$, it is evaluated by multiplying state $S_{i}S_{j}$ by probability $S_{h,i}^{n}S_{h,j}^{n}$. It should be noted that $S_{x}$ includes an indeterminate parameter $c$ that characterizes the system as a Diophantine problem, as shown in Eq.~(\ref{eq:statevecresultcasea}) in Section~\ref{sec:targetsysfeature}. We also mentioned that \HL{$c=l$} is required for normalization when we refer to $S_x$ as the probability. \HL{However}, it should be emphasized that parameter $c$ is unspecified when referring to $S_{x}$ as a system state. Accordingly, it is necessary to describe the parameter $c$ when mentioning an arbitrary state $S_x$. This is significant because there is a certain possibility that the parameter $c$ can be a major reason for the observed nonlinear phenomena. Because $S_{h,i}^{n}$ is already normalized, as in Eq.~(\ref{eq:statevecapprox}), we can express $S_{i}$ by $S_{h,i}^{n}$ as $S_{i} \approx \bar{c}S_{h, i}^{n}$, \HL{where $\bar{c}=c/l$}. The mean $\langle S_{i} S_{j} \rangle$ can be expressed as $(\bar{c}S_{h,i}^{n} S_{h,j}^{n})^2$. 
In this stage, the covariance can be represented as $\sigma_{ij} = (\bar{c}S_{h,i}^{n} S_{h,j}^{n})^2 - \{(l+1)M\}^{-2}$.

\HL{Recall} that the dependence of $I_{h}$ on $M$ in case (C) was observed to approach that in case (A) as parameter $\alpha$ increased. Specifically, the surges of $I_h$ in a small area of $M$ were observed to decrease as $\alpha$ increased, as shown in Fig.~\ref{fig:Figure_Dependence_IhonM}; agents were equally distributed among secondary nodes as the parameter $\alpha$ increased. This \HL{corresponds to} the fact that the agent distribution in a stationary state gets closer to the uniform distribution as $\alpha$ increases. We can say that the correlations among different secondary nodes become \HL{negligible} when agents are kept equally distributed among secondary nodes, compared to the case of a biased agent distribution. This is further explained as follows: when the amount of usage in different secondary nodes is equal among secondary nodes, the second terms in the bracket of the numerator in Eq.~(\ref{eq:considerconges}) for all \HL{secondary} nodes have approximately equal values. \HL{Thus}, the difference in their congestion \HL{rates} hardly contributes to the agents' decision-making to choose destinations. \HL{In brief}, the covariance decreases as $\alpha$ increases. To reflect this characteristic, we introduced a phenomenological scale parameter $C_d$ that controls the intensity of $\sigma_{ij}$. Consequently, we obtained the following \HL{expression} for variance $\sigma_{ij}$:
\begin{eqnarray}
\sigma_{ij} \approx C_d\biggl\{ \bigl(\bar{c}~S_{h,i}^{n} S_{h,j}^{n}\bigr)^2 - \frac{1}{(l+1)^2M^2}\biggr\}. \label{eq:estimsigmaC}
\end{eqnarray}

\bibliographystyle{h-physrev3}
\bibliography{reference}

\end{document}